\documentclass{aastex6}

\begin{document}


\title{New Candidate Planetary Nebulae in Galactic Globular Clusters from the VVV Survey}
\footnote{Based  on  observations  taken  within  the  ESO  programmes
  179.B-2002 and  298.D-5048}

\author{
Dante Minniti\altaffilmark{1,2,3},  
\and Bruno Dias\altaffilmark{2,1}, 
\and Mat\'ias G\'omez\altaffilmark{1},
\and Tali Palma\altaffilmark{4},
\and Joyce B. Pullen\altaffilmark{2}
}

\altaffiltext{1}{Depto. de Ciencias F\'isicas, Facultad de Ciencias Exactas, Universidad Andres Bello, Av. Fernandez Concha 700, Las Condes, Santiago, Chile.}
\altaffiltext{2}{Instituto Milenio de Astrof\'isica, Santiago, Chile.}
\altaffiltext{3}{Vatican Observatory, V00120 Vatican City State, Italy.}
\altaffiltext{4}{Observatorio Astron\'omico de C\'ordoba, Universidad Nacional de C\'ordoba, Laprida 854, C\'ordoba, Argentina.}


\begin{abstract}
Only four globular cluster planetary nebulae (GCPN) are known so far in the Milky Way.
About 50 new globular clusters have been recently discovered towards the Galactic bulge.
We present a search for planetary nebulae within 3 arcmin of the new globular clusters, revealing the identification of new candidate GCPN.
These possible associations are 
PN SB 2  with the GC Minni 06,
PN G354.9-02.8  with the GC Minni 11,
PN G356.8-03.6  with the GC Minni 28, and
PN Pe 2-11  with the GC Minni 31.
We discard PN H 2-14 located well within the projected tidal radius of the new globular cluster FSR1758 because they have different measured radial velocities.
These are interesting objects that need follow-up observations (especially radial velocities) in order to confirm membership, and to measure their physical properties in detail.
If confirmed, this would double the total number of Galactic GCPN.
\end{abstract}




\section{Introduction} 
\label{sec:intro}


Planetary nebulae (PN) that are confirmed members of (GCs) are very rare. 
Asymptotic giant branch (AGB) and white dwarf stellar masses in old globular clusters typically are  $\sim 0.5M_{\odot}$ which is less than the minimum mass needed  to form a PN (Kalirai et al. 2009).
There are only four known GCPN in the Milky Way GC system, namely 
K 648 in M15, IRAS 18333-2357 in M22 (NGC6656), JaFu1 in Pal6, and JaFu2 in NGC6441 (Jacoby et al. 1997). 
This number of only four Galactic GCPN has remained constant for the past two decades, and
it is smaller than expected from stellar evolution theory, because the fuel consumption theorem of Renzini \& Buzzoni (1986)  predicts four times as many (16 GCPN instead of 4). A solution to this open question is the possibility that single stars in GCs are unable to produce PN, and binary interactions are needed (see Bond 2015).

PN that are members of star clusters are important. As one example, Majaess et al. (2007) studied the possible membership for thirteen PNe in Milky Way open clusters, concluding that seven of these are real pairs.
More recently, Moni-Bidin et al. (2014) reviewed the state of the art on this subject, and investigated other potential Galactic planetary nebula--open cluster pairs.
They studied four candidates PN VBe 3 in NGC 5999, PN HeFa 1 in NGC 6067, PN He 2-86  in NGC4463, and PN NGC 2452 in NGC 2453.
However, the first two of these were discarded as cluster members based on radial velocities.

As another example, elliptical galaxies contain numerous PN and also GCs, both tracers used to study their distances, distributions, kinematics, and metallicities, especially in their outer halo regions, where observations of the integrated light are difficult (see e.g. Kissler-Patig 2000, Arnaboldi 2012, Goudfrooij et al. 2013, Richtler et al. 2014, Mendez 2017, and references therein).
However, confirmed extragalactic GCPN are also rare, with only a few cases known.
Minniti \& Rejkuba (2002) found one candidate in the globular cluster G169 belonging to the
peculiar E/S0 galaxy NGC5128 (Centaurus A).  This object was discovered and spectroscopically confirmed using the  Magellan I Baade Telescope.
Also, Jacoby et al. (2013) confirmed spectroscopically five candidates GCPN in the Andromeda galaxy (M31), out of 270 objects sampled. 

There have also been a few PN confirmed members of a couple of dwarf galaxies satellites of the Milky Way.
For example, there are two confirmed PN in the Fornax dwarf spheroidal Galaxy.
The first one was discovered and confirmed spectroscopically by  Danziger et al. (1978).
The other PN is associated with the GC H5 of the Fornax dSph galaxy (Larsen 2008).
Also, presently there are four confirmed PN members of the Sgr dwarf galaxy: Wray 16-423, StWr 2-21, PN BoBn 1, and He 2-436 (Zijlstra \& Walsh 1996, Walsh et al. 1997, Zijlstra et al. 2006), but none of those Sgr dwarf PN are associated with a GC.

The handful GCPN known are, however, precious objects in the sense that they serve as  distance calibrators, and are relevant for the study of more distant populations.
In this paper we present new GCPN candidates in the Milky Way bulge, with the goal of encouraging follow-up observations of these objects.
Section 2 describes the search for new GCPN candidates using the recently available databases.
In Section 3 we discuss the candidate GCPN found in detail.
Finally, the conclusions of this work are summarized in Section 4.

\section{Searching for New GCPN Candidates} 
\label{sec:sec2}

Dozens of new GC candidates have been recently discovered  in the central regions of the MW, thanks mainly to the near-IR photometric surveys 
(e.g. 2MASS -- Skrutskie et al. 2005, Glimpse -- Benjamin et al. 2003, VVV -- Minniti et al. 2010).
Recently Bica et al. (2019) published the most complete  and up to date list of Galactic GC candidates, including a number of new GC discoveries by his team and others, {
that are not included in the classic GC catalog of Harris (1996, 2010 edition).}
In particular, we consider a total of 50 new Galactic GCs already published by different groups as follows:
1 GC from Longmore et al. (2011),
1 GC from Minniti et al. (2011),
1 GC from Ortolani et al. (2012),
1 GC from Minniti et al. (2017a),
2 GCs from Borissova et al. (2014), 
22 GCs from Minniti et al. (2017b) and Piatti (2018),
5 GCs from Camargo (2018), 
2 GCs from Ryu \& Lee (2018), 
2 GCs from Cantat-Gaudin (2018),
3 GCs from Camargo \& Minniti (2019), and
10 GCs  from Palma et al. (2019).

In parallel, a number of new PN have been found in the Galactic bulge, largely by the Macquarie-AAO-Strasbourg surveys (MASH I and II; Parker et al. 2006, Miszalski et al. 2008).
Here we use the Simbad Database at the Centre des Donnees Stellaires at Strasbourg in order to query a variety of existing bulge PN catalogues
(e.g. Acker et al. 1991, 1992,  Durand et al. 1998, Beaulieu et al. 1999, Kohoutek 2001, 2002, Parker et al. 2006, Miszalski et al. 2008, etc.) available at 
{\it http://http://simbad.u-strasbg.fr/}.

We have searched for known PN within 3' of the center of these 50 new GC candidates, finding a few matches.
The 3' search radius is arbitrary, but sufficiently small to minimize random  coincidences and also to match the small size of the new GC candidates 
(with the exception of FSR1758 which is much larger, as discussed in Section 3). 


{We cannot rule out a chance superposition for the GCPN candidates presented here. Indeed, as both the PN and GC samples increase, so does the probability of chance superposition as Frew et al. (2016) throughly discuss. We therefore encourage further attention to these objects since they are -if true GCPN- of utmost astrophysical importance, as they are only objects where we can fix the mass and metallicity of both the PN and its progenitor.}


{Taking into account that there were 4 GCPN known in about 150 Galactic GCs studied so far, we can also compute the expected number of new candidates.                                          
Jacoby et al. (1997) point out that the known GCPN associations are preferentially found in clusters with large fractions of x-ray binaries. Therefore, 
depending on the (unknown) relaxation times of these new clusters compared to those of the previously known GCs, a simple scaling may not work. 
For a rough estimate, there are 50 new candidate GCs explored here, and they are low luminosity GCs in comparison with the already known clusters.                                                      
Therefore the expectation is to find only 1 or 2 GCPN in this new sample, where we found 4 new candidates. However, these are small number statistics, and                                        
the possibility of chance alignment between a GC and a field PN cannot be completely discarded.                                                                                                   
Considering them individually, the GCs Minni 11, and 31 are located in very crowded regions  with  the highest stellar density (Figures 1 to 3),                                                  
so the possibility of a chance alignment with a field PN is higher for these clusters.}

The new candidate PN located within 3' of the GC centers are listed in Table 1. 
We give the cluster IDs and positions (RA and DEC coordinates), the VVV tile IDs, the GC distances from the Sun, the number of RRLyrae variable stars within 3 arcmin of the GC centers, the PN IDs, the radial velocities of the PN, and the PN separations from the GC centers in arcsec.

The Wide-field Infrared Survey Explorer (WISE) satellite observed the whole sky in mid-IR wavelengths (Wright et al. 2010).
PN are particularly conspicuous in the near and mid-IR, especially dusty PN (e.g. Anderson et al. 2012).
Figure 1 shows the WISE mid-IR fields for the candidate GCPN. 
In some cases the PN are clearly seen in these composite mid-IR images.

Figure 2 shows the corresponding optical images centered on the GCPN candidates from the DECAPS survey (Schlafly et al. 2018), and 
Figure 3 shows the VVV near-IR images, made using the $JHKs$ filters for all the candidate GCPN, except for PN H 2-14 in FSR1758 that 
is not shown because it is discarded as discussed below. These figures illustrate that the appearance of these GCPN is different depending on the wavelength. 
These images are complementary to those found in the MASH database, obtained using narrow band filters (Parker et al. 2006, Miszalski et al. 2008).

\section{Discussion} 
\label{sec:sec3}
The available data on the new candidate PN located within 3 arcmin of the GC centers are listed in Table 2. 
We give the PN IDs and positions (in Equatorial and Galactic coordinates), the PN radial velocities, and the PN sizes in arcsec.
The new GCPN candidates are discussed in turn below.

\subsection{PN SB 2  in the GC Minni 06 Field} 
Minni 06 is a metal-poor GC located in the bulge at a distance  $D=8.4 \pm 1.5$ kpc (Minniti et al. 2017a).
This distance measurement was confirmed by Piatti (2018), who measured a cluster size $r_h=2.9$ pc, equivalent to 1.2'. 

PN SB 2 (a.k.a. PN G000.5-05.3) has a radial velocity $V_r=+3$ km/s (Beaulieu et al. 1999) that is consistent with either disk or bulge kinematics.
{Further observations, like radial velocity measurements for the GC or Balmer decrements of the PN would be extremely useful in confirming the true nature of this association.}

\subsection{PN G354.9-02.8  in the GC Minni 11 Field} 
Minni 11 is a metal-rich GC located in a dense bulge region, with high extinction ($A_K=0.33$).
Its distance is uncertain, ranging from  $D=5.9 \pm 1.5$ kpc (Minniti et al. 2017a) to $D=11.7 \pm 2.2$ kpc (Piatti 2018), and it is also a small cluster with angular size $r_h=1.2'$ (equivalent to $\sim 2-4$ pc at these distances). 

PN G354.9-02.8 is not a well studied PN, although it is a clear PN from the optical and near-IR images, with a size $r=10.2"$ that is typical for PN located at the distance of the Galactic bulge (Ruffle et al. 2004).
{There are no radial velocity measurements of the GC and the PN in the literature, both are highly needed to confirm/discard membership.}

\subsection{PN G356.8-03.6  in the GC Minni 28 Field} 
Minni 28 is classified as a metal-poor GC, and appears to be located in the far side of the bulge, at $D=10.1 \pm 1.5$ kpc (Palma et al. 2019).

PN G356.8-03.6 (a.k.a. PHR J1752-3330) was classified as a possible PN by the MASH survey (Parker et al. 2006), with a size $r=25.8"$  (equivalent to $\sim 1.2$ pc at that distance).
We confirm that this is a true PN, that has a clear double shell structure seen in the available optical and near-IR images.
The diffuse outer shell is rather large for the distance of the GC, and more consistent with a nearby object. 
However, the size of the inner shell is smaller $r=7.5"$, and $r=3.0"$ as measured from the DECAPS optical images (Schlafly et al. 2018), and the 
VVV near-IR images, respectively, which does not rule this object out as member of the GC Minni 28. 

There are no radial velocity measurements of this PN available in the literature, but there is spectroscopy confirming that the central star is an object with Wolf-Rayet emission lines, classified as a [WC11] type (DePew et al. 2011). This star also has measured Gaia DR2 (Gaia Collaboration 2018)
proper motions: $\mu_l = -5.215 \pm  0.095, ~ \mu_b= -9.136 \pm  0.070$ mas/yr, that are consistent with kinematics of both the disk and the bulge, but the parallax is uncertain.

The VVV near-IR images clearly show the central point source, and Weidmann et al. (2013) report near-IR photometric measurements for this object:
$J =13.282 \pm 0.052$, $H=13.012 \pm 0.049$, and  $K=12.825 \pm 0.052$ mag. These magnitudes are consistent with a PN located within the bulge.

\subsection{PN Pe 2-11  in the GC Minni 31 Field} 
The GC Minni 31 is also metal-poor, and located in the bulge at  $D=9.1 \pm 1.5$ kpc (Palma et al. 2019). 

PN Pe 2-11 (a.k.a. PN G002.5-01.7) has a large radial velocity $V_r=+155$ km/s, that is consistent with the bulge kinematics (Durand et al. 1998), ruling out a foreground disk object.
{However, a radial velocity for the GC or measurements of the Balmer decrements for the PN are necessary to assess membership.}

Also Weidmann et al. 2013 report near-IR photometry from the VVV survey for this object:
$J =14.189 \pm 0.075$, and  $H=13.197 \pm 0.069$ mag. This photometry and the small measured size $r=4.8"$ are consistent with a PN located at the distance of the GC.
In spite of its small size, this object exhibits a clear asymmetry in the optical and near-IR images, resembling other well known bipolar PN.

\subsection{The case of PN H 2-14 in the Field of  FSR1758} 

FSR1758 is a large GC recently discovered in the Galactic bulge (Cantat-Gaudin et al. 2018, Barba et al. 2019).
This GC contains at least a dozen RRLyrae,  and it is now a relatively well studied metal-poor GC with $[Fe/H]=-1.5$ dex (Villanova et al. 2019).

The PN H 2-14 (a.k.a. PN G349.2-03.5, Beaulieu et al. 1999),  is located in the vicinity of this new GC.
Figure 2 shows the WISE mid-IR finding charts for the field of this candidate GCPN. 
This PN is located at 13.4 arcmin away from the GC center, but the size of this cluster is estimated to be very large ($R_t=47$ arcmin).
Therefore, this PN is well within the projected tidal radius of FSR1758, and this is why it is also included in the list of possible candidate Galactic GCPN (Table 1).
The  size of PN H 2-14  is also consistent with the sizes of  PN at the distance of the bulge as  measured by Ruffle et al. (2004).

However, the radial velocity of PN H 2-14 measured by Beaulieu et al. (1999) is $V_r=-12$ km/s, is more consistent with either disk or bulge kinematics, 
and much lower than the mean systemic velocity of FSR1758. 
This GC has a confirmed retrograde orbit, with $V_r = +227$ km/s, based on the Gaia RV measurement of Simpson (2019), and  $V_r=226.8 \pm 1.6$ km/s, 
based on high dispersion spectroscopy of 9 red giant star members by Villanova et al. (2019).

In consequence, the association of  PN H 2-14 with FSR1758 seems unlikely. Nonetheless, further observations of this PN are warranted in order to accurately measure its physical parameters (chemical composition, proper motion, distance, size, etc.).

The association of the sample PN with the GCs is important to augment the sample of PN with well known distances. 
In general the PN sizes are consistent with the sizes of PN located at the distance of the new bulge GCs, as measured by Ruffle et al. (2004). 
But this consistency argument is not enough to confirm membership to the GCs.  
Kinematic confirmation of these GCPN candidates is needed, as done for example for the PN members of the Sgr dwarf galaxy, that were confirmed using radial velocities (Zijlstra \& Walsh 1996, Zijlstra et al. 2006).

The association of the sample PN with the GCs is also important for studies of chemical composition of the stellar populations (including their measurement systematics), and the understanding of the chemical evolution of the Galactic bulge, as discussed by Chiappini et al. (2009). 
The new GCPN would help to tie in the samples of new GCs and new PN being discovered, that would feed in turn massive future spectroscopic surveys like 4MOST (Chiappini et al. 2019). 

\section{Conclusions} 
\label{sec:sec4}

By matching the positions of fifty newly discovered GCs  in the Milky Way bulge to the existing PN catalogs we found five new GCPN candidates
(Table 1). 
Out of these, we argue that PN H 2-14 is probably not a GCPN associated with the GC FSR1758, because its membership appears to be ruled out by the existing radial velocities. 
On the basis of the existing multiwavelength data explored here, the following objects remain as excellent GCPN candidates:
PN SB 2  in the GC Minni 06,
PN G354.9-02.8  in the GC Minni 11,
PN G356.8-03.6  in the GC Minni 28, and
PN Pe 2-11  in the GC Minni 31.
Their individual membership has to be confirmed with additional observations, {like radial velocities, accurate proper motions and parallaxes or Balmer decrements.}

There is now the possibility of making progress in a field that has remained stagnant for quite some time, perhaps doubling  the number of  known GCPN in the Milky Way.
However, much remains to be learned, and aside from the identification, the purpose of this paper is to encourage further observations of these new candidates.
These new objects  open the way for interesting follow-up studies, such as the measurement of kinematics, chemical compositions and luminosities. 
Also, these objects are bright enough to have accurate parallax measurements within the future Gaia data release, scheduled for late 2020, and for follow-up spectroscopy with the 4MOST survey at the ESO VISTA telescope (Chiappini et al. 2019).
Finally,  studying the GC-PN connection could lead to a better understanding of GC stellar content and evolution.

\acknowledgments
We gratefully acknowledge data from the ESO Public Survey program ID 179.B-2002 taken with the VISTA telescope, and products from the Cambridge Astronomical Survey Unit (CASU). This publication makes use of data products from the WISE satellite, which is a joint project of the University of California, Los Angeles, and the Jet Propulsion Laboratory/California Institute of Technology, funded by the National Aeronautics and Space Administration. This research has made use of NASA’s Astrophysics Data System Bibliographic Services and the SIMBAD database operated at CDS, Strasbourg, France.  Support is provided by the BASAL Center for Astrophysics and Associated Technologies (CATA) through grant AFB170002, and the Ministry for the Economy, Development and Tourism, Programa Iniciativa Cientifica Milenio grant IC120009, awarded to the Millennium Institute of Astrophysics (MAS). D.M. and M. G. acknowledge support from FONDECYT Regular grant No. 1170121. We would also like to thank very much the useful comments by Daniel Majaess.\\


\begin{deluxetable}{lccrrrrccc}
\tablecaption{PN in New Milky Way Bulge GC Candidates \label{tab1}}
\tablehead{
\colhead{Cluster ID} & \colhead{RA2000} & \colhead{DEC2000} & \colhead{VVV tile}  & \colhead{Distance}  & \colhead{$N_{RRL}$} & \colhead{PN ID}  & \colhead{separation$^a$} 
}
\startdata
Minni 06   &  18 08 22 & -31 06 18 & b264&8.4 kpc$^b$   &   0  & PN SB 2                & 165" \\
Minni 11   &  17 44 33 & -34 43 22 & b288&5.9 kpc$^b$   &   2  & PN G354.9-02.8    & 143" \\
Minni 28   &  17 52 32 & -33 30 00 & b289&10.1 kpc$^c$ &  4  & PN G356.8-03.6    &  39" \\
Minni 31   &  17 58 36 & -27 38 21 & b307& 9.1 kpc$^c$  &   5  & PN Pe 2-11            & 105" \\
FSR 1758&  17 31 12 & -39 48 30 & e683&11.5 kpc$^d$ &  12   & PN H 2-14             & 804" \\
\enddata
\tablenotetext{a}{Separation between the PN and the GC center in arcsec.}
\tablenotetext{b}{GC distance from Minniti  al. (2017a).}
\tablenotetext{c}{GC distance from Palma  al. (2019).}
\tablenotetext{d}{GC distance from Barba  al. (2019).}
\end{deluxetable}


\begin{deluxetable}{lccrrrrccc}
\tablecaption{Available Data on New Bulge GCPN Candidates \label{tab1}}
\tablehead{
\colhead{PN ID} & \colhead{RA2000} & \colhead{DEC2000} & \colhead{l(deg)} & \colhead{b(deg)} & \colhead{$V_r$(km/s)} & \colhead{Size(arcsec)} 
}
\startdata
PN SB 2              &18 08 34.70 &-31 06 52.0  &  0.59830 &-5.39180      &   3.0  &  9.5$^b$, 20.0$^c$  \\ %
PN G354.9-02.8  &17 44 43.50 &-34 44 26.0  &354.94540 &-2.86340    &   ---   & 10.2$^a$, 8.5$^c$  \\ %
PN G356.8-03.6  &17 52 29.22 &-33 30 04.2  &356.83678 &-3.60549    &   ---   &  25.8$^a$, 3.0$^b$, 7.5$^c$  \\ 
PN Pe 2-11          &17 58 31.27 &-27 37 05.8  &  2.58188 &-1.77190      & 155.0 & 4.8$^b$, 4.8$^c$  \\ 
PN H 2-14            &17 32 20.09 &-39 51 25.6  &349.29469 &-3.50161   & -12.0 &  15.0$^c$ \\ %
\enddata
\tablenotetext{a}{Diameters of the PN listed in the MASH catalogs (Parker et al. 2006, Mizsalski et al. 2008).}
\tablenotetext{b}{Diameters of the PN estimated from the VVV near-IR images.}
\tablenotetext{c}{Diameters of the PN estimated from the DECAPS optical images (Schlafly et al. 2018).}
\end{deluxetable}

\begin{figure}[h!]
\begin{center}
\includegraphics[scale=0.36,angle=0]{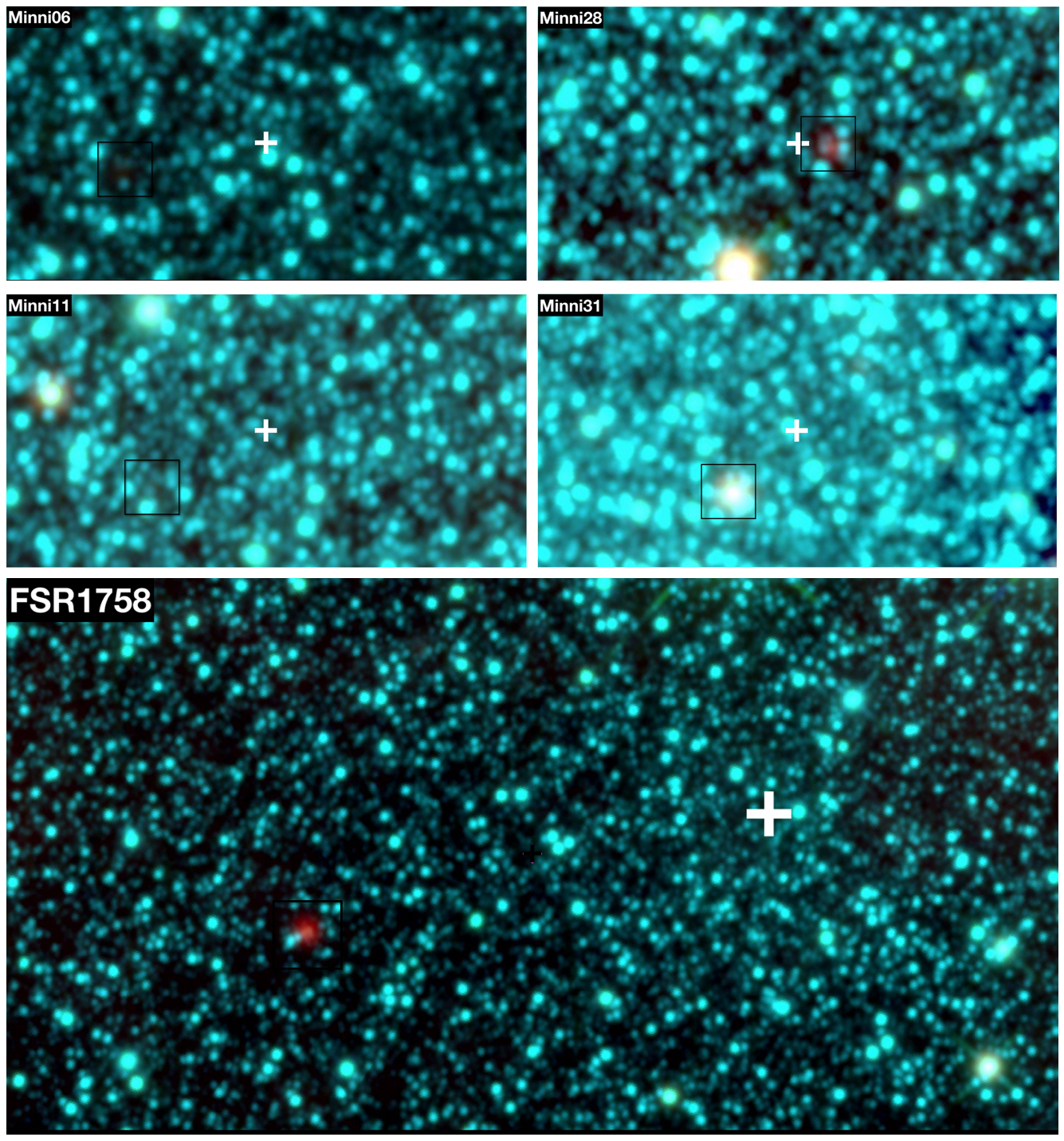}
\caption{Top: WISE satellite (Wright et al. 2010) mid-IR finding charts centered on the candidate GCs (white crosses) that show the location of the known PN (large black squares).
The fields of view are $8' \times 4'$, and are oriented with North on top and East to the left.
Bottom: WISE satellite mid-IR finding chart for the PN H 2-14 (reddestmost object), located 804" away from the center of the GC FSR1758 (white cross).
The field of view is $30' \times 15'$, and is oriented with North to the top and East to the left.
\label{fig:1}}
\end{center}
\end{figure}

\begin{figure}[h!]
\begin{center}
\includegraphics[scale=0.20,angle=0]{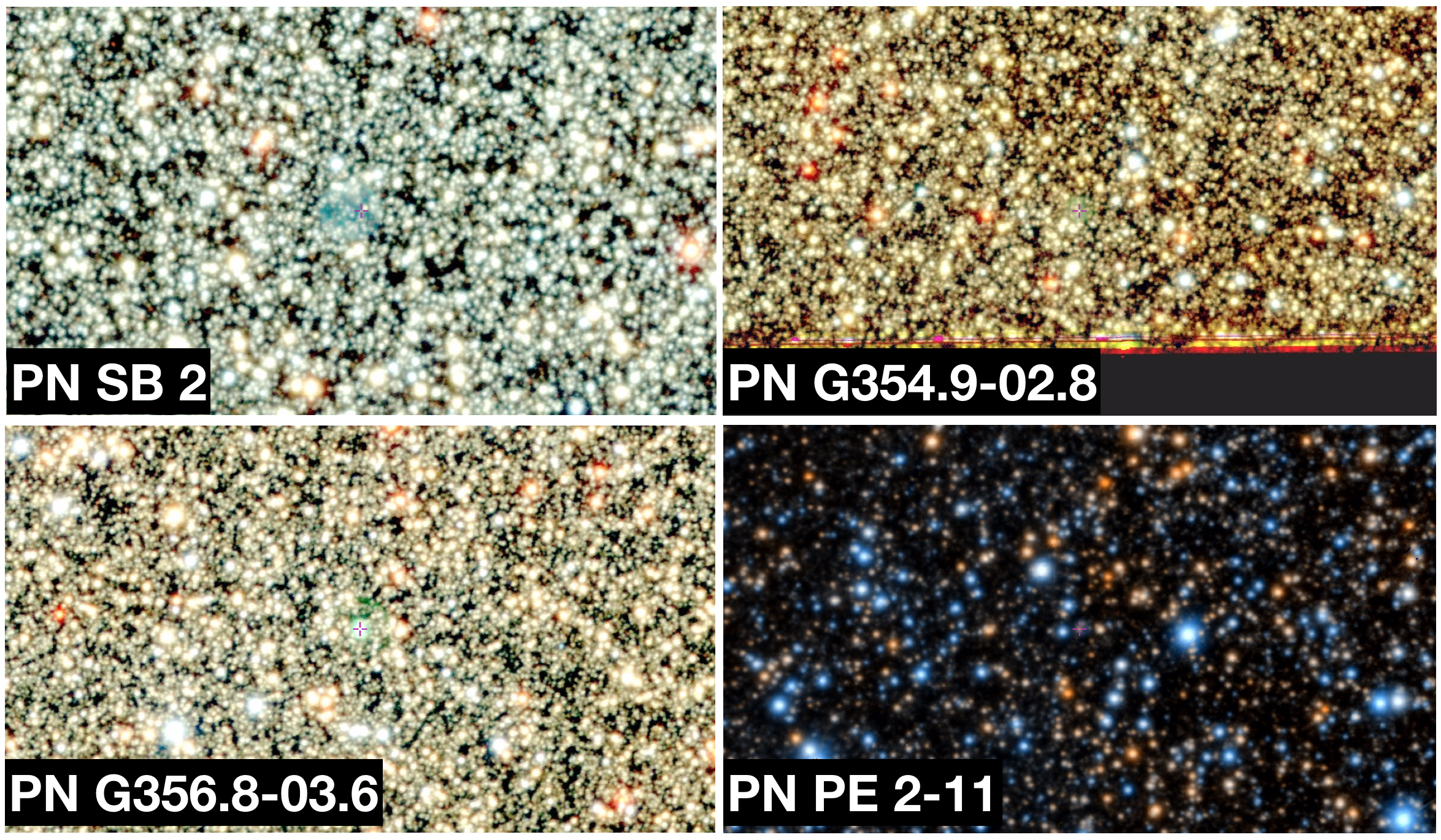}
\caption{DECAPS optical finding charts for the candidate GCPN listed in Table 2 (except for PN Pe 2-11 for which we show the PanStarrs composite image available from the Simbad Database).
The fields of view is $8' \times 4'$, and are oriented with North to the top and East to the left.
\label{fig:1}}
\end{center}
\end{figure}

\begin{figure}[h!]
\begin{center}
\includegraphics[scale=0.18,angle=0]{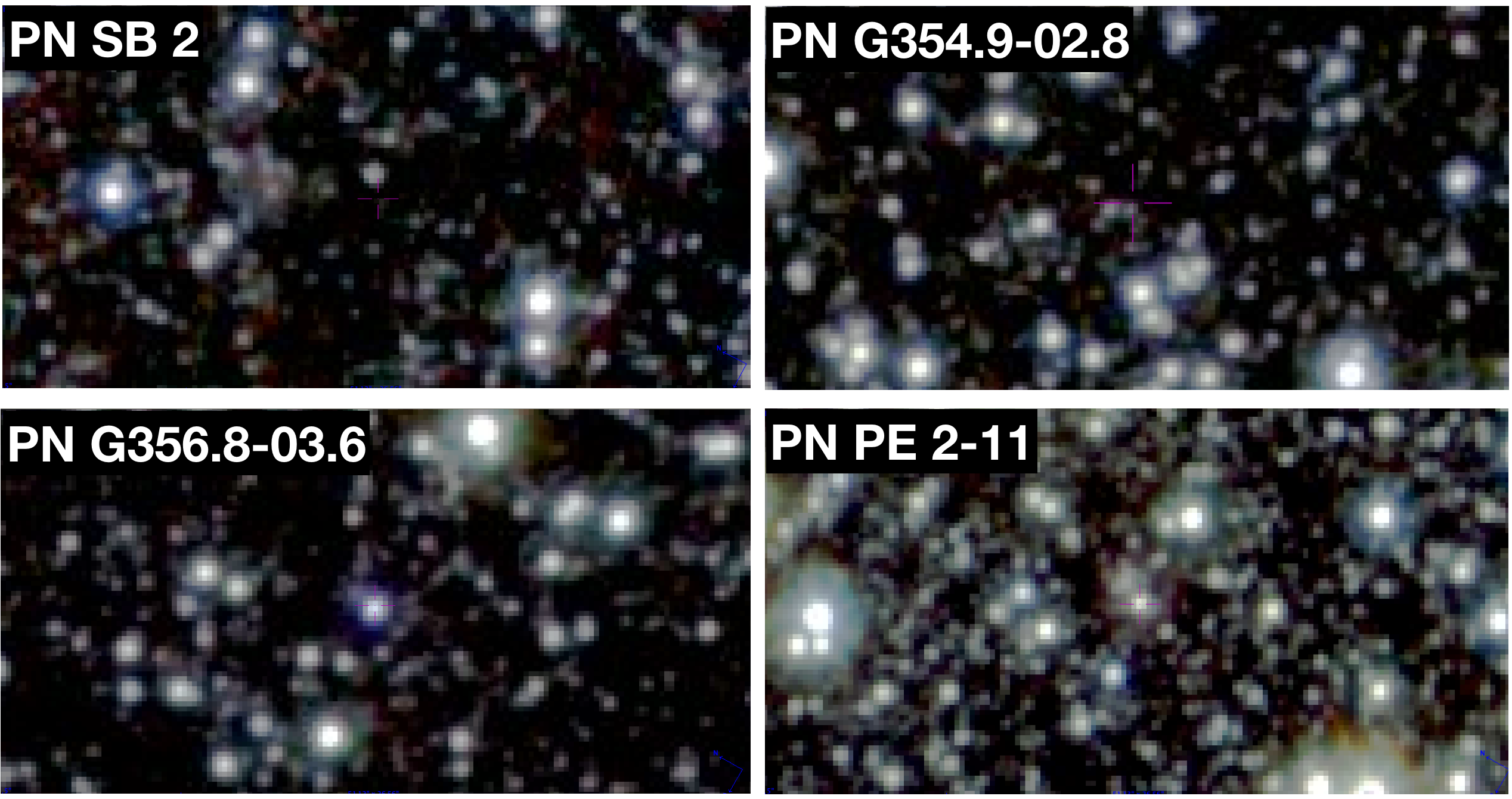}
\caption{Zoomed VVV near-IR fields for the candidate GCPN listed in Table 2.
The fields of view is $50" \times 25"$, and are oriented along Galactic coordinates increasing to the top and to the right.
\label{fig:1}}
\end{center}
\end{figure}

\pagebreak

\end{document}